\def\apjl{ Astrophys. J. Lett.}
\def\apj{ Astrophys. J.}

\def\aap {Astron. Astrophys.}
\def\mnras{Mon. Not. R. Astron. Soc.}
\def\pasp{Publ.  ASP}

\def\aapr{Astron. Astrophys. Rev.}

\documentclass[symmetry,article,accept,oneauthor,pdftex]{Definitions/mdpi} 

\firstpage{1} 
\pubvolume{xx}
\issuenum{1}
\articlenumber{5}
\pubyear{2020}
\copyrightyear{2020}
\history{Received: 30 September 2020; Accepted: 24 November 2020; Published: 27 November 2020}
\updates{yes} 

\Title{Strongly Lensed Supernovae in Well-Studied Galaxy Clusters with the Vera C. Rubin Observatory}

\Author{Tanja Petrushevska 
 \orcidA}

\AuthorNames{Tanja Petrushevska}

\address[1]{%
Centre for Astrophysics and Cosmology, University of Nova Gorica, Vipavska 11c, 5270 Ajdov\v{s}\u{c}ina, Slovenia; tanja.petrushevska@ung.si}

\corres{Correspondence: tanja.petrushevska@ung.si}

\abstract{Strong lensing by galaxy clusters can be used to significantly expand the survey reach, thus allowing observation of magnified high-redshift supernovae that otherwise would remain undetected. Strong lensing can also provide multiple images of the galaxies that lie behind the clusters. Detection of strongly lensed Type Ia supernovae (SNe Ia) is especially useful because of their standardizable brightness, as they can be used to improve either cluster lensing models or independent measurements of cosmological parameters. The cosmological parameter, the Hubble constant, is of particular interest given the discrepancy regarding its value from measurements with different approaches. Here, we explore the feasibility of the Vera C. Rubin Observatory Legacy Survey of Space and Time (LSST) of detecting strongly lensed SNe in the field of five galaxy clusters (Abell 1689 and Hubble Frontier Fields clusters) that have well-studied lensing models. Considering the 88 systems composed of 268 individual multiple images in the five cluster fields, we find that the LSST will be sensitive to SNe~Ia (SNe~IIP)  exploding in 41 (23) galaxy images. The range of redshift of these galaxies is between $1.01 < z < 3.05$. During its 10 years of operation, LSST is expected to detect  $0.2\pm0.1$ SN~Ia and $0.9\pm0.3$ core collapse SNe. However, as LSST will observe many more massive galaxy clusters, it is likely that the expectations are higher. We stress the importance of having an additional observing  program for photometric and spectroscopic follow-up of the strongly lensed SNe detected by LSST.
}
\usepackage{gensymb}
\usepackage{amssymb}
\begin{document}
	\section{Introduction}
	Supernovae (SNe) have proved to be invaluable tools for various astrophysical and cosmological applications.  The strong gravitational lensing effect is another powerful tool; it occurs when a foreground mass distribution is located along the line of sight to a background source. The observer can see multiple images of the background source appearing around the foreground lens (see, e.g., in~\cite{2019RPPh...82l6901O} for a review).  As the images have taken different paths through space before reaching us, the time differences between the images are sensitive to the expansion rate of the universe, parametrized by the Hubble constant, \emph{H$_0$} \cite{Taubenberger_2019}. These time delays are sensitive to other cosmological parameters, albeit to a much lesser extent \cite{2016A&ARv..24...11T}. Using time delay of strongly lensed variable sources for measuring  \emph{H$_0$} is also known as the Refsdal method  as it was suggested by the author more than 50 years ago  \cite{1964MNRAS.128..307R}. The time delays depend also on the lensing potential; therefore, strongly lensed SNe can be used for both cosmological and cluster lens studies \cite{2002A&A...393...25G,2009ApJ...706...45C,Third,2018ApJ...860...94G}. 
	
	In recent years, a discrepancy at least at the $4.4\sigma$ level is emerging between the value of \emph{H$_0$} inferred from the indirect measurement from the cosmic microwave background \cite{Planck2018} and the value derived from the local distance scale composed of classical Cepheid variable stars and SNe Type Ia (SNe~Ia) \cite{2016ApJ...826...56R,2018ApJ...855..136R,Riess2019}. The~Refsdal method provides a third independent alternative to measure \emph{H$_0$}, and as strongly lensed SNe are rare events, it has been used on quasars (see, e.g., in \cite{2010ApJ...711..201S,2016A&ARv..24...11T}).  The H0LiCOW collaboration measured \emph{H$_0$} to 2.4\% accuracy, combining six lensed quasars, each with 6--10\% precision \cite{Wong_2020}. However, quasar light curves are stochastic and unpredictable; therefore, they require long observations spanning over years and even more than a decade (see, e.g., in \cite{2010ApJ...711..201S}). Relative to quasars, lightcurves of SNe~Ia form a homogeneous group (see, e.g., in \cite{2011ARNPS..61..251G,2019ApJ...876..107P}). Knowing the absolute brightness of SNe~Ia allows to estimate the absolute magnification of SNe~Ia, and therefore to break the so-called mass-sheet degeneracy of gravitational lenses~\cite{2001ApJ...556L..71H}. Thus, they could be used to put constraints on the lensing potential, if a cosmological model is assumed (see, e.g., in \cite{2014MNRAS.440.2742N,2014ApJ...786....9P,2015ApJ...811...70R}). 
	
	Systematic searches for SNe in background galaxies behind clusters have been performed with both space- and 
	ground-based instruments \cite{2002MNRAS.332...37G,Second,2014ApJ...786....9P,2015ApJ...811...70R,Petrushevska2016,Petrushevska2018a}. Very large Hubble Space telescope (\textit{HST}) programs with more than thousand orbits have been targeting galaxy clusters, to improve cluster lensing models and search for SNe \cite{2015ApJ...812..114T,2017ApJ...837...97L}. Remarkably, they resulted in the discovery of the first multiply-imaged SN (dubbed ``SN Refsdal'') behind a galaxy cluster  \cite{Kelly2015, Kelly2016} and it was most probably  a core-collapse (CC) type of explosion~\cite{Kelly2016}. Several teams predicted the reappearance of SN Refsdal almost a year later, which allowed testing of their lens models \cite{2016ApJ...822...78G, 2016ApJ...819L...8K}. By measuring the time delays of SN Refsdal and having a high-quality strong lensing model of the galaxy cluster, it was shown that it is possible to measure \emph{H$_0$} with 6\% total uncertainty~\cite{2020ApJ...898...87G}. Dedicated ground-based searches for lensed SNe behind galaxy clusters have been performed using near-infrared instruments at the Very Large Telescope \cite{2009A&A...507...61S, Second, Petrushevska2016, Petrushevska2018a}. Most notably, they reported the discovery of one of the most distant CC~SN ever found, at redshift $z = 1.703$ with a lensing magnification factor of $4.3 \pm 0.3$ \cite{2011ApJ...742L...7A}. Furthermore,  thanks to the power of the lensing cluster, it was possible to estimate the volumetric CC~SN rates for $0.4\leq z< 2.9$ and compare it with the predictions from cosmic star formation history \cite{Petrushevska2016}.
	
	The first resolved multiply-imaged SN Ia, iPTF16geu, with four resolved images arranged symmetrically around the lens galaxy,  was magnified by a galaxy lens \cite{2017Sci...356..291G}. Before the discovery of iPTF16geu, only one other strongly lensed SN~Ia by galaxy lens was known, PS1-10afx, but only few years later it was realized that it was a SN~Ia magnified $\sim30$ times by the foreground galaxy~\cite{2014Sci...344..396Q}. PS1-10afx most probably had multiple images, which could not be resolved with the available data~\cite{2013ApJ...768L..20Q,2014Sci...344..396Q}. The multiple images of iPTF16geu were well resolved by \textit{HST} and ground-based images with adaptive optics, but the estimated time delays were about half a day, which is difficult to measure with SNe~Ia light curves~\cite{Dhawan_2019}. Cluster lensing timescales are much longer, typically at the order of months and years \cite{Third, Petrushevska2016,Petrushevska2018a,Petrushevska2018b}. Therefore, the microlensing effects, i.e., the stellar-scale lensing which is important at micro-arcsecond scales and causes fluctuations on timescales of weeks to months \cite{Foxley_Marrable_2018}, become subdominant. These two considerations could make cluster lens time delays measurement potentially more feasible, in the cases where the lens potential is well studied, so the predicted time delays have small uncertainties \cite{2018ApJ...860...94G}.
	
	In this work, we focus on prospects of observing SNe which are strongly lensed by well-studied galaxy clusters with the upcoming Vera C. Rubin Observatory (Rubin Observatory, hereafter). The Rubin Observatory  will scan the transient sky through its Legacy Survey for Space and Time (LSST) with a 8.4~m telescope, targeting at least 18,000 deg$^2$ of the southern hemisphere with a field of view of 9.6~deg$^2$~\cite{collaboration2017sciencedriven}. Its duration is set to at least ten years and it will use set of six broad photometric bands, $ugrizy$.  The~LSST observing strategy has not been finalized \cite{collaboration2017sciencedriven}, which means  that distribution of observations to a field within a year and the distribution between filters have not been decided. Choosing an observing strategy is an important and challenging task as the main science goals of the LSST are quite broad, ranging from constraining dark energy to exploring the Solar System, the Milky Way, and the transient optical sky. There will be a main ``wide-fast-deep'' (WFD) survey which will consume $\sim80\%$ of the whole survey time, covering the equatorial declination range $-62\degree < \delta < +2 \degree$, and excluding the central portion of the Galactic plane. In the WFD survey, two visits in either the same or different filters are acquired each night, to allow identification of moving objects and rapidly varying transients. These~pairs of visits are repeated every three to four nights throughout the period the field is visible in each year. Currently, the official LSST observation is a pair of 15 s exposures, which helps to reject cosmic rays, but there is a possibility that it could be switched a single 30 s exposure because of a potential gain of 7\% efficiency and improved image quality \cite{2018arXiv181200515L}.
		Here, we consider the baseline observing strategy\footnote{\url{https://cadence-hackathon.readthedocs.io/en/})}which uses 2 $\times$ 15 s exposures, simulated with the LSST Operations Simulator (OpSim) \cite{collaboration2017sciencedriven}, based on a model of the observatory (including telescope) and historical data of observational conditions \cite{Biswas_2020}. We use the latest baseline OpSim simulation released in August 2020 \footnote{\url{http://astro-lsst-01.astro.washington.edu:8080/?runId=16}}. 
	
	\section{Estimating the Expected Number of Strongly Lensed Supernovae Observable by the  Rubin Observatory}
	We consider the six galaxy clusters from the {\it Hubble Frontier Fields} (HFF)  \cite{2017ApJ...837...97L} program with the \textit{HST}. These are Abell~2744, MACS~J0416.1-2403, MACS~J0717.5+3745, MACS~J1149.5+2223, Abell~S1063, and Abell~370. We add to this list the well-studied galaxy cluster Abell~1689. The reason for this choice is the existence of reliable lensing models for these clusters, thus well constrained magnification maps, based on good quality data, such as deep \textit{HST} images and Multi-Unit Spectroscopic Explorer (MUSE) spectroscopy. Using recently published lensing models of the HFF clusters  \cite{2017MNRAS.469.3946L, 2018ApJ...855....4K,2018MNRAS.473..663M}, the predicted magnifications and time delays between the images were presented in \cite{Petrushevska2018a, Petrushevska2018b}. In the same studies, by using multi-band \textit{HST} photometry, the global properties of the multiply-imaged galaxies were inferred. These properties are the stellar mass, star formation rate (SFR), and SN rate, which depend on the luminosity of the galaxy corrected for the predicted magnification. This means that the inferred galaxy properties depend on the lensing model. 
	
\begin{figure}[H]
		\centering
		\includegraphics[width=14cm]{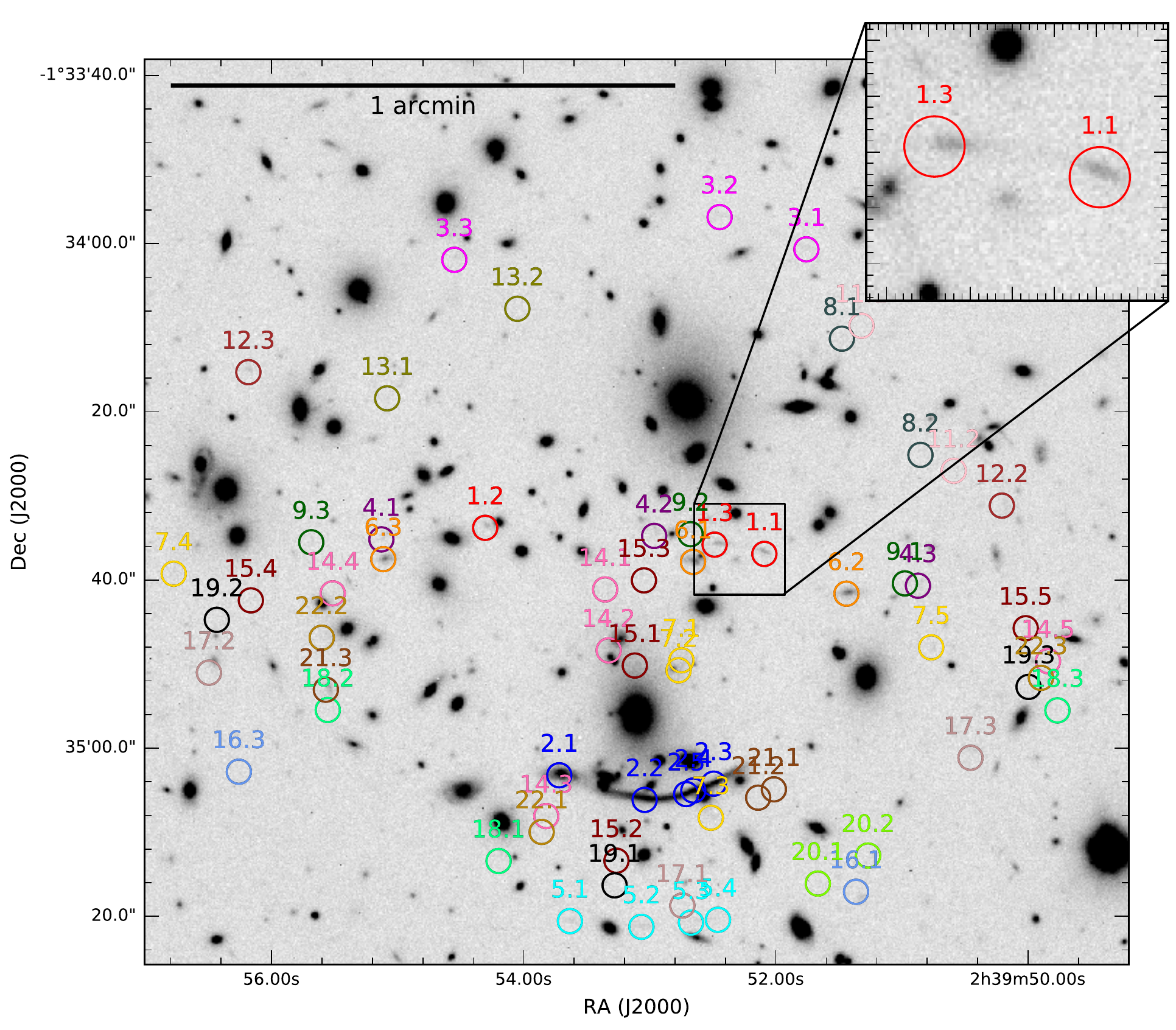}
		\caption{Galaxy cluster Abell~370 overplotted with the positions of all the 21 background galaxies with 67 multiple images with expected time delays presented in \cite{Petrushevska2018a}. The images that belong to the same system are shown with the same color. The predicted magnifications for these galaxies from the lensing model are approximately $\sim$1--4 magnitudes. As an example, a  zoomed-in view of multiply-imaged galaxies 1.3 and 1.1 is shown, which are magnified by $2.81\pm0.08$ and $2.80\pm0.07$ mag, respectively. Adopted from the work in \cite{Petrushevska2018a}.}
		\label{A370}
	\end{figure}

\begin{figure}[H]
		\centering
		\includegraphics[width=12cm]{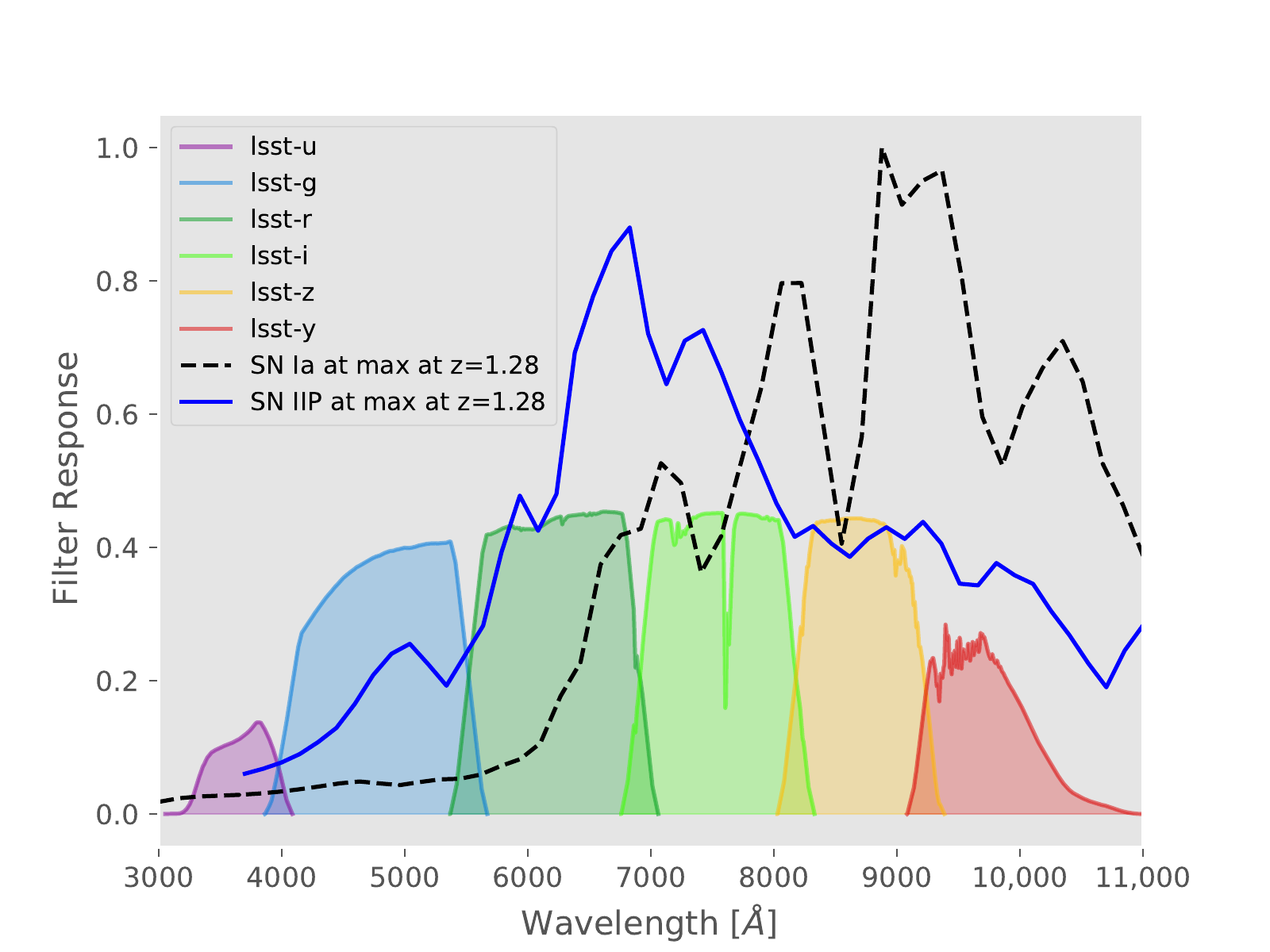}
		\caption{The filter response system that will be on the Rubin Observatory overplotted with a typical spectra of SN Ia and SN IIP. The redshift of the SN spectra is at 1.28, as the examples from Figure~\ref{lightcurves} where multiply-imaged galaxies in Abell~370 are shown  at which the survey will be sensitive, considering the image depth and the magnification from the galaxy cluster.} 
		\label{fig:filter}
	\end{figure}

	From these seven galaxy clusters, two (MACS J1149.5+2223 and MACS J0717.5+3745) are not visible from the latitude of the Rubin Observatory. The final list is given in the Table \ref{table:clusters} together with number of the background galaxies,  number of multiple images that these galaxies have, and their redshifts. An example of one galaxy cluster with the background galaxies and its 67 multiple images are shown in Figure~\ref{A370}. We only consider the multiply imaged galaxies that have a spectroscopic redshift. Considering the redshifts of the SNe from Table~\ref{table:clusters}, the most important filters are at the longest wavelengths: $i$, $z$, and $y$. The reason for this is because most of the light of nearby SNe is in the optical bands, so their light is redshifted to the longer wavelengths when higher redshifts are considered. An example of typical SN~Ia and SN~IIP at redshift that can be observed in the galaxy cluster fields considered here is given in Figure~\ref{fig:filter}, together with the LSST filter response system. Most of the pairs of images that have time delays that are shorter than ten years (see column 3 of Table \ref{table:results1} from the work in \cite{Petrushevska2018a} and column 3 of Table \ref{table:results2} from the work in \cite{Petrushevska2018b}), which is the planned duration of LSST.  In Table~\ref{table:clusters}, we also include the number of visits to the clusters that the LSST will dedicate in  ten years. In the baseline observing strategy considered here, there is no large difference between the number of visits per filter or per cluster because LSST is designed to observe at least 825 times each field over the duration of the WFD survey. However, the image depth differs between filters; it is the highest in the $i$ and the lowest in the $y$ band. The  average $5\sigma$ image depth of the selected cluster fields in $i$ band is $\sim 23.4$, $\sim22.7$ in $z$, and $\sim 22.0$ in $y$ band. 
	\begin{table}[H]
		\centering
			\caption{The galaxy clusters considered in this work.  The number of unique galaxies behind the cluster is given in column 2, and the number of their multiple images of these galaxies in column 3. The redshift range  of these galaxies is given in column 4. Columns 5--7 show the numbers of observations that the cluster field will have in the 10 years of the LSST, divided by filter $izy$. }
			\label{table:clusters}
			\begin{tabular}{lcccccc}
				\toprule
				\textbf{Cluster} & \boldmath{N$_{systems}$} & \boldmath{N$_{images}$} & \boldmath{\emph{z}$_{min-max}$} & \textbf{Observed}  & \textbf{Observed} & \textbf{Observed}\\ 
				&&& &\textbf{10 yrs, Band} \boldmath{$i$} &\textbf{10 yrs, Band} \boldmath{$z$} &\textbf{10 yrs, Band} \boldmath{$y$} \\
				\midrule
				Abell 1689  & 18 & 51 & 1.15--3.4 & 188 & 167 & 175\\ 
				Abell 370  & 21 & 67 & 0.73--5.75 & 194 & 174 & 174 \\ 
				Abell 2744  & 12 & 40 & 1.03--3.98 & 198 & 181 & 170\\ 
				Abell S1063  & 14 & 42 & 1.03--3.71 & 203 & 180 & 189\\ 
				MACS J0416.1-2403  & 23 & 68 & 1.01--3.87 & 200 & 176 & 189\\ 
				\midrule
				Total & 88 & 268 \\
				\bottomrule
			\end{tabular}
	\end{table}
	
	We use the procedures outlined in \cite{Petrushevska2016} to calculate the sensitivity of LSST to detect  strongly lensed SNe in the multiply lensed background galaxies of the five galaxy clusters. First, we estimate the control time for each galaxy image, $T_i$, which indicates the amount of time the survey is sensitive of detecting a SN candidate \cite{1938ApJ....88..529Z}. The control time above the detection threshold, is a function of the SN light curve, absolute intrinsic SN brightness, the image depth, dates of observation, filter, extinction by dust and the lensing magnification. The peak V band brightness and its one standard deviation is assumed as $19.3\pm0.3$ for SNe~Ia, while for CC~SNe we use the values compiled by the authors of \cite{Vincenzi_2019} in their Table~\ref{table:clusters}.  The probability distribution of the absolute intrinsic brightness is assumed to be Gaussian. As in \cite{Petrushevska2016, Petrushevska2018a}, the values for the color excess were drawn from a positive Gaussian distribution with a mean $E(B-V)=0.15$ and $\sigma_E=0.02$ justified by extinction studies of high-redshift galaxies behind Abell 1689 \cite{2014ApJ...780..143A}.  The dates of observation,  filter, and image depth are taken from the  simulated survey strategy generated with~OpSim.
	
\begin{figure}[H]
		\centering
		\includegraphics[width=9cm]{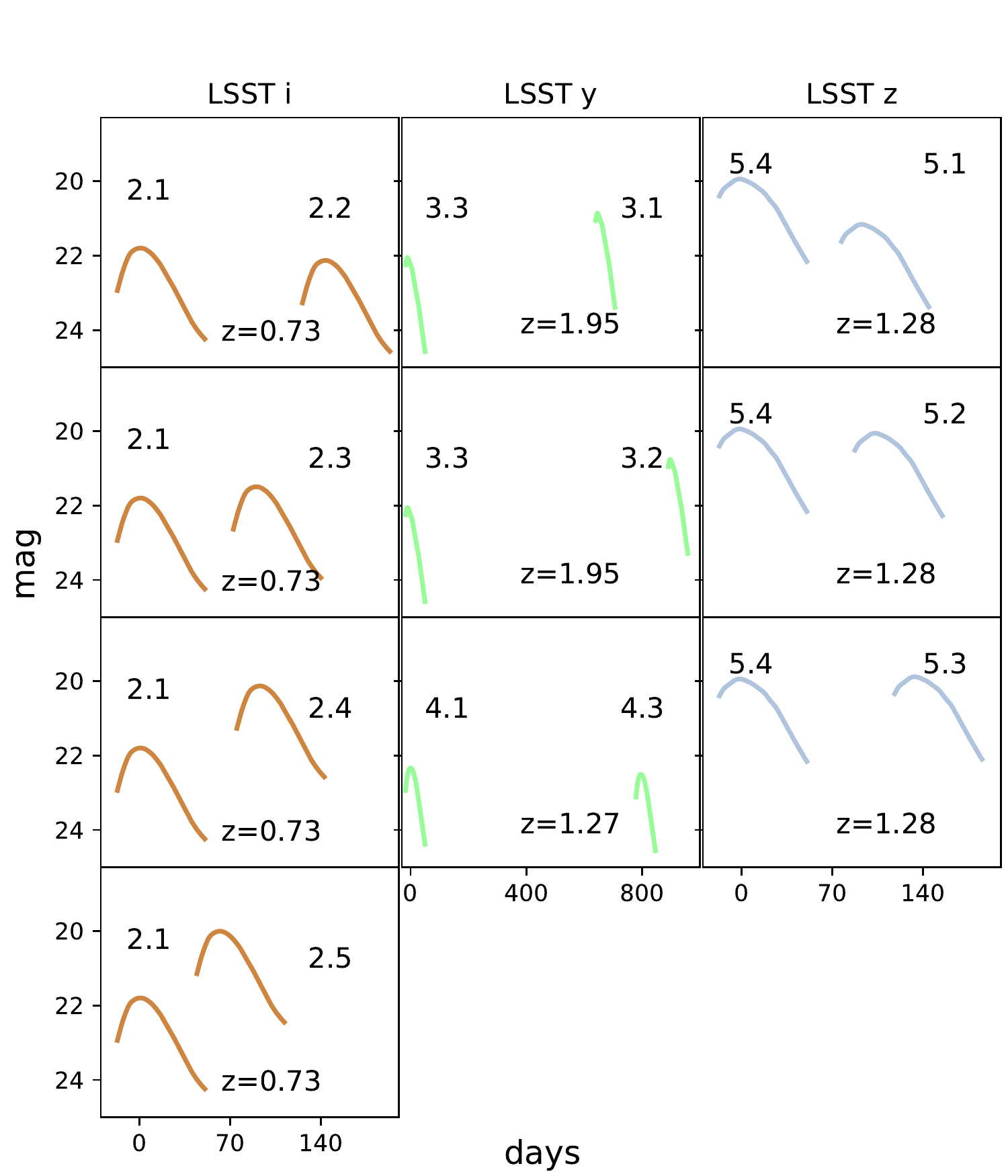}
		\caption{Examples of simulated light curves of SNe Ia for pairs of images of strongly lensed galaxies behind Abell~370. The first column are in the LSST $i$ band, the second in the $y$, while the third is in $z$ band. If a SN explodes in one of the galaxies, the explosion should appear in the corresponding images, thus allowing for scheduling observations accordingly. Without the magnification from the galaxy cluster, the SNe~Ia exploding in these galaxies would hardly be observable with the Rubin Observatory.}
		\label{lightcurves}
	\end{figure}

	LSST can detect a SN in those galaxies that have positive control time. Few examples of simulated SN~Ia light curves that could be observed with LSST are shown in Figure~\ref{lightcurves}.  The synthetic light curves in the observer filters for redshift $z$ were obtained by applying cross-filter k-corrections \citep{1996PASP..108..190K}. Furthermore, to obtain the ``observed'' magnitude in the chosen LSST filter,  the lensing magnification from the galaxy cluster is also taken into consideration. For example, the multiple images (labeled 5.1, 5.2, 5.3, and 5.4 in the last column in Figure~\ref{lightcurves}) of galaxy source 5 at redshift z $ =1.28$, are  magnified by $2.99\pm0.08$, $4.09\pm0.09$, $4.27\pm0.12$, and $4.21\pm0.13$ mag, respectively (Table~\ref{table:results1} from the work in \cite{Petrushevska2018a}). With this magnification, typical SNe are detectable with the LSST, even at this high redshift. Furthermore, the estimated time delays between 5.4 and the other three images is $94\pm4$, $104\pm8$, and $135\pm7$ days,  respectively (Table~\ref{table:results1} from the work in \cite{Petrushevska2018a}). The length of these time delays is within the duration of LSST, so if a SN explodes in this galaxy, there is a possibility that it will be detected in the other images during the survey,  as the median revisit of the same field is 1 to 8 days, depending on the filter. However,  there are gaps of observations up to 200--300 days, due to, for example, visibility issues. Therefore, it is not guaranteed that LSST will observe that specific field when the other SN images appear, and observations with another photometric and spectroscopic program should be scheduled when the reappearance is expected.
	\unskip    
	\begin{table}[H]
		\caption{{\small Sensitivity of LSST to SNe~Ia exploding in the the multiply-imaged galaxies in HFF galaxy clusters + Abell 1689 fields. Columns 2 (4) show the number of galaxy images in which LSST could detect a SN~Ia (SN IIP) explosion. In columns 3 and 5 the largest redshift of these galaxies is reported . }}
		\label{table:results1} \centering \begin{tabular}{lcccccc} \toprule  \textbf{Cluster} &   \boldmath{N$_{gal  Ia}^c$} & \boldmath{$z_{max Ia}$}  & \boldmath{N$_{gal  IIP}^c$} & \boldmath{$z_{max IIP}$}   \\
			\midrule	
			Abell 1689 & 8 & 1.83 &  4 & 3.05 \\ 
			Abell 370 & 17 & 1.95 & 14 & 2.75 \\ 
			Abell 2744 & 5 & 1.04 &  0 & / \\ 
			AS1063 & 9 & 1.26 & 2 & 1.26 \\ 
			MACSJ0416 & 2  & 1.01 & 3 &  2.28 \\ 
			\midrule 
			Total & 41 &  & 23 &  \\ 
			\bottomrule

		\end{tabular}

	\end{table}

	\begin{table}[H]
		\caption{{\small Expectations for the number of lensed SNe in the multiply-imaged galaxies that could be detected behind the HFF galaxy clusters + Abell 1689 in the 10 years of the Rubin Observatory. The errors in the N$_{Ia}$ and N$_{CC}$ originate from the propagated uncertainty in the SFR. }}
		\label{table:results2} \centering \begin{tabular}{lccc} \toprule  \textbf{Cluster} &   \boldmath{\textbf{N$_{Ia}$}} &  \boldmath{\textbf{N$_{CC}$}} \\
			&\textbf{10 yrs} &\textbf{10 yrs} \\
			\midrule	
			Abell 1689 & $0.13\pm0.10$ & $0.24 \pm 0.17$\\ 
			Abell 370 & $0.04\pm0.01$ & $0.41\pm0.24$  \\ 
			Abell 2744 & $0.0012\pm0.0001$ & $0.02\pm0.01$\\ 
			Abell S1063 & $0.014\pm0.01$ &  $0.14\pm0.10$\\ 
			MACS J0416.1-2403 & $0.014\pm0.004$ & $0.10\pm0.05$ \\ 
			\midrule 
			Total & $0.20\pm0.10$ & $0.91\pm0.31$\\ 
			\bottomrule

		\end{tabular}

	\end{table}

	Second, we calculate the expected number of SNe in each galaxy $N_i$, by multiplying the SN rate $R_i$ and the control time $T_i$,
	\begin{equation}
		N_i=R_i\cdot T_i,
	\end{equation}
	where $i$ indicates the individual galaxy. As the control time is dependent on the SN type, it is computed separately for each SN type. The total CC control time is obtained by weighting the contributions for the CC SN types. We use the values for the relative CC~SN fractions from \cite{Vincenzi_2019}, compiled in their Table~1. The~expected SN~Ia rate in a galaxy, $R_i$, depends on the SFR and the stellar mass through an empirical relation \cite{2005ApJ...629L..85S}, while the CC~SN rate scales with the SFR. We used the SFR, stellar masses  and SN rates estimates from the works in \cite{Petrushevska2018a, Petrushevska2018b} for the HFF clusters and from the work in \cite{Third} for Abell 1689. To obtain the total expected number of SNe~Ia and CC SNe over all the systems, we sum the expectations over the individual galaxies. 
	\section{Results}
	First, considering the five well-studied galaxy clusters and their background strongly lensed galaxies with multiple images, we find that the LSST will be sensitive to SNe~Ia exploding in 41 galaxies, while for the most common CC~SNe, SNe~IIP, the number is lower, 23. The main reason for this is that SNe~IIP are simply fainter. The galaxies for which LSST is sensitive of detecting SNe have redshift between $0.73 < z < 1.95$ for SNe~Ia, while for CC~SNe $0.73 < z < 3.05$. SNe~IIP are visible to LSST at higher redshifts because they emit much more in the rest-frame UV, so at high redshift that region is shifted and it is still visible with the LSST filters  at the longest wavelengths (see Figure~\ref{fig:filter}). As SNe~Ia are not bright at  rest-frame UV, they are practically not observable at $z\gtrsim2$ with an optical survey. The sensitivity of LSST for each galaxy cluster is shown in Table~\ref{table:results1}. 
	
	Second, we show the estimated number of expected  strongly lensed SNe in the selected galaxy cluster fields in 10 years of the Rubin Observatory, presented in Table \ref{table:results2}. The observability of a SN in the multiply-imaged galaxies is sensitive to the SN properties such as light curve, absolute intrinsic SN brightness; galaxy properties such as redshift, star formation rate, extinction in the line of site and magnification of the galaxy, but also on the survey parameters such as image depth, cadence and filter. Considered the baseline observing strategy of the 10 years of  operation, LSST is expected to detect $0.2\pm0.1$ SN~Ia and $0.9\pm0.3$ CC~SN  in the five cluster fields. However, since we have considered only SNe in multiply-imaged galaxies with spectroscopic redshift of five galaxy clusters, this is a lower limit. As~there are more galaxies with photometric redshift only, it is likely that the expectations are higher. We note that, as the LSST official observing strategy might be adjusted in the future, our forecast will also change.



	
	

	
	\section{Discussion}
	The Rubin Observatory is not optimal for the discovery of strongly lensed SNe in galaxy cluster fields, given that it is a ground-based observatory with filters which do not extend to the near-infrared and infrared, where most of the light is emitted of SNe at redshifts considered here (see Figure~\ref{fig:filter}). The problem is especially accentuated for SNe~Ia. Therefore, for this purpose, a space-based observatory with extended near-infrared and infrared filters such as the upcoming James Webb Space Telescope (JWST \cite{2018ConPh..59..251K})  or the Nancy Grace Roman Space Telescope \cite{2015arXiv150303757S} would be more appropriate. By considering only four visits of one hour which can be taken in one year in the broad near-infrared filter F150W to the clusters considered here, the JWST/NIRCam instrument is expected to find $0.19\pm0.07$ SNe~Ia \cite{Petrushevska2018b}. When CC~SNe are considered, the expected number is much higher, $1.7\pm0.5$. SN rates at high redshifts are dominated by CC~SNe, so it is likely that the strongly lensed SNe by galaxy clusters will by dominated by the CC type \cite{Petrushevska2016,Petrushevska2018b}.  Obtaining observing time for a dedicated multi-year search with JWST relatively small field of view of $2.2' \times 2.2'$ will most-likely be competitive with high over-subscription rates. On the other hand, LSST observations of the galaxy cluster fields will be part of the nominal program and the data can be searched for strongly lensed SNe immediately. However, the Rubin Observatory will serve to detect the strongly lensed SNe, but it is better to have additional follow-up by other photometric and spectroscopic instruments for detailed study of the lensed SNe and  their time delays. 
	
	One way to improve LSST sensitivity to high-redshift SNe is to  obtain better image depth in the reddest bands ($izy$) through co-adding images from visits closely separated in time.  Improving the image depth means that the survey becomes sensitive to detecting SNe in more multiply-imaged galaxies. This can be done through a dedicated program which will focus on monitoring the galaxy cluster fields. A similar strategy is currently implemented by the Zwicky Transient Facility \cite{2014htu..conf...27B} for searching for strongly lensed SNe~Ia by lens galaxies \cite{2017ApJ...834L...5G, 2018ApJ...864...91S, Goldstein_2019}.
	
As we noted in the previous section, the results presented in this work are lower limits, as we have only considered five galaxy clusters and the multiply-imaged galaxies in the field with spectroscopic redshift. It is expected that LSST will observe $\sim 70$ galaxy clusters with Einstein radii larger than $\theta_E >20 "$ \cite{2009arXiv0912.0201L}. More systems behind galaxy clusters will be measured with spectroscopic campaigns with instruments such as MUSE at the Very Large Telescope (see, e.g., in \cite{2016A&A...590A..14B,2017MNRAS.469.3946L,2018MNRAS.473..663M}), which not only will increase the number of multiply-imaged galaxies with spectroscopic redshifts in the clusters considered here, but will also allow to have more galaxy clusters with high quality lensing models.
	
	For the purpose of inferring \emph{H$_0$} with high precision, it would be most useful to measure time delays of SNe~Ia, which have a well-established relationship between their optical peak luminosity and rate of brightness decline, which makes them excellent standardizable candles. However, detecting strongly lensed and magnified CC~SNe, which are much more common at high redshifts \cite{Strolger_2015}, is also beneficial for several reasons. Despite the fact that SN Refsdal is a CC type of explosion, thanks to the high-quality strong lensing model of the galaxy cluster, it was shown that it is possible to measure \emph{H$_0$} with 6\% total uncertainty \cite{2020ApJ...898...87G}. Furthermore, the discovery and the  photometric and spectroscopic follow-up data that can be obtained could also help understand CC~SNe at high redshifts. This would be interesting, as CC~SNe are on average intrinsically fainter than SNe~Ia and often embedded in dusty environments, making their studies at $z>0.4$ challenging (see, e.g., in \cite{2016MNRAS.457.1107H}). Therefore, the magnification from galaxy clusters can provide an advance in redshift and can even allow the constraints of the CC~SN rate at high redshifts~\cite{Petrushevska2016}.  As both the area and the flux of background galaxies are magnified, it also allows for accurately studying the host environment, which would not be possible in unlensed scenario  \cite{2011ApJ...742L...7A}. Therefore, even if the LSST detects the more common case, strongly lensed CC~SNe with significant magnification, but no multiple images, they can still be useful for the means of studying SNe and their host environment at unprecedented redshifts.
	\section{Conclusions}
	As there is a current tension regarding the \emph{H$_0$} value, one of the priorities of the community is to provide an independent measure  of  \emph{H$_0$}, which can be done by monitoring time delays of cluster lensed SNe. There is a further utility of magnified SNe by clusters. They can  be used as test of galaxy cluster lens models, or in the cases of well-known lensing models, there is the opportunity to study SNe and their host environment in the early universe.
	
	In a previous work, we estimated the magnifications and the time delays of the multiply-imaged galaxies behind HFF clusters based on high-quality lens models by the authors of  \cite{2017MNRAS.469.3946L,2018MNRAS.473..663M,2018ApJ...855....4K}. We also inferred their properties such as SFRs, stellar masses, and SN rates by using very deep \textit{HST} photometry. Here, we explored the feasibility of detecting strongly lensed SNe with the Rubin Observatory behind these galaxy clusters. We found that in the 10 years of LSST in the five galaxy cluster fields, the expectations are $0.2\pm0.1$ SN~Ia and $0.9\pm0.3$ CC~SN, though these are likely lower limits, as LSST will observe dozens more massive galaxy clusters. Given the redshift range of $0.73<z<5.75$, the ground-based Rubin Observatory filter set is not optimal for detecting SNe in these galaxies. Next generation space telescopes will be much more appropriate for this task. Nevertheless, massive clusters as those considered in this study will work as precious gravitational telescopes providing concrete possibility for LSST to observe strongly SNe at high redshift.
	
	\vspace{6pt} 
	
	\funding{This research was funded by the Slovenian Research Agency (grants I0-0033, P1-0031, J1-8136 and Z1-1853).}
	
	
	\conflictsofinterest{The author declares no conflict of interest.} 
	
	\abbreviations{The following abbreviations are used in this manuscript:\\
		
		\noindent 
		\begin{tabular}{@{}ll}
		 CC SN&	Core collapse supernova  \\
		\emph{H$_0$}&	Hubble constant    \\
		HFF	& Hubble Frontier Fields  \\
		 \textit{HST}&	Hubble Space telescope  \\
		 JWST&	James Webb Space Telescope  \\
		 LSST&	Legacy Survey for Space and Time \\
		 SFR	&Star formation rate \\
			 SN &Supernova  \\
		 SN~Ia&	Supernova Type Ia  \\
		 Rubin Observatory 	&Vera C. Rubin Observatory 
	\end{tabular}}
	
	\appendixtitles{no} 
	\unskip
	
	
	\reftitle{References}
	
	

	
	
\end{document}